\documentclass[doublecol]{epl2}
\usepackage{graphicx}
\usepackage{bm}

\newcommand{\be}{\begin{equation}}
\newcommand{\ee}{\end{equation}}
\newcommand{\beqn}{\begin{eqnarray}}
\newcommand{\eeqn}{\end{eqnarray}}

\title{Random transverse-field Ising chain with long-range interactions}

\author{R. Juh\'asz\inst{1}\thanks{E-mail: \email{juhasz.robert@wigner.mta.hu}} \and I. A. Kov\'acs\inst{1,2}\thanks{E-mail: \email{kovacs.istvan@wigner.mta.hu}} \and F. Igl\'oi\inst{1,2}\thanks{E-mail: \email{igloi.ferenc@wigner.mta.hu}}}

\shortauthor{R. Juh\'asz \etal}

\institute{
  \inst{1} Wigner Research Centre, Institute for Solid State Physics and Optics,
H-1525 Budapest, P.O.Box 49, Hungary, EU\\
  \inst{2} Institute of Theoretical Physics, Szeged University, H-6720 Szeged, Hungary, EU
}
\date{\today}

\abstract{
We study the low-energy properties of the long-range random transverse-field Ising chain with ferromagnetic interactions decaying as a power $\alpha$ of the distance. Using variants of the strong-disorder renormalization group method, the critical behavior is found to be controlled by a strong-disorder fixed point with a {\it finite} dynamical exponent $z_c=\alpha$. Approaching the critical point, the correlation length diverges exponentially. In the critical point, the magnetization shows an $\alpha$-independent logarithmic finite-size scaling and the entanglement entropy satisfies the area law. These observations are argued to hold for other systems with long-range interactions, even in higher dimensions.
}


\begin{document}
\maketitle
\textbf{Introduction.} - It is well known that long-range (LR) interactions decaying in $d$ dimension as a power of the distance, $J(r)\sim r^{-(d+\sigma)}$,
are able to modify the universality class of systems with short-range (SR) interactions for sufficiently
small decay exponents, $0<\sigma<\sigma_L$. 
Furthermore, LR couplings may lead to spontaneous ordering and critical behavior
even in dimensions below the lower critical dimension of the corresponding SR system. 
In general, the study of the critical properties of LR models is
technically very demanding both theoretically and numerically. 
In the Ising (and n-vector) models, the critical behavior is non-universal\cite{fisher_me}, i.e. the critical exponents are
$\sigma$-dependent for
$\sigma_U < \sigma < \sigma_L$, while the transition is mean-field-like for $0<\sigma < \sigma_U$. However, for $d \ge 2$
the actual value of $\sigma_L$ and the functional form of the varying critical exponents are still under debate\cite{sak,bloete,picco,parisi}.
Classical LR models have been studied in the presence of quenched disorder, too, such as the random-field Ising model\cite{rfim} or
Ising spin glasses\cite{sg,young,katzberger,cecile1} and similar $\sigma$-dependent critical regimes have been found.

In the case of quantum phase transitions, which take place at zero temperature the effect of LR interactions has been considered, motivated by recent progress of experimental studies of trapped ions in optical lattices simulating spin
models\cite{friedenauer,kim,islam,britton,islam1}.
Most of the theoretical studies are for systems with dipolar interactions\cite{porras,deng,hauke,peter,nebendahl,wall},
i.e. for $d+\sigma=3$, but there are also investigations,
in which the decay exponent $\sigma$ is a free parameter\cite{cannas,dutta,dalmonte,koffel,hauke1}. In this field of research
we mention a Monte Carlo study of the quantum Ising chain with
ohmic dissipation\cite{werner}, which corresponds to a LR quantum Ising chain with $\sigma=1$\cite{sandvik}. The critical behavior of this model is found to be anisotropic,
characterized by a dynamical exponent $z \approx 2$ and a correlation-length exponent $\nu \approx 0.63$.

In the presence of quenched disorder, quantum LR systems have not been systematically studied yet. In this Letter, we aim at filling this gap and consider a prototype of such systems, 
the random transverse-field Ising model defined by the Hamiltonian
\be
{\cal H} =
-\sum_{i\neq j} \frac{b_{ij}}{r_{ij}^\alpha} \sigma_i^x \sigma_{j}^x-\sum_{i} h_i \sigma_i^z\;,
\label{eq:H}
\ee
with $\alpha\equiv d+\sigma$ in terms of the Pauli matrices $\sigma_i^{x,z}$ at lattice site $i$. 
Here $r_{ij}$ denotes the distance between site $i$ and $j$,  and the parameters $b_{ij}$ and transverse fields $h_i$ are independent,
positive, quenched random variables drawn from some distributions $p_0(b)$ and $q_0(h)$, respectively. 
A closely related real system is the compound
$\rm{LiHo}_x\rm{Y}_{1-x}\rm{F}_4$ placed in a transverse field\cite{experiment}, apart from that, here, the transverse field also
induces a random longitudinal field via the off-diagonal terms of the dipolar interaction\cite{rf}.
Since, at least for $d=\sigma=1$, the fixed point of the pure system is unstable against quenched disorder\cite{harris},
the critical behavior of the model in Eq.(\ref{eq:H}) is expected to be controlled by a new LR random fixed point.

Here we consider mainly the one-dimensional model and study its critical properties by variants of the so called
strong disorder renormalization group (SDRG) method\cite{im}. 
As a first step, we investigate the problem numerically by an efficient
algorithm of the SDRG method. After this, we analyse the typical renormalization steps and
introduce a simplified scheme, the so called
primary model, which is expected to contain the essential ingredients of the SDRG procedure and has the same asymptotic
scaling properties as the original model. Then, we analytically solve the SDRG equations for the primary model
and the obtained critical properties are compared with the numerical SDRG results for the original model. Subsequent to this, some of the scaling results are explained in terms of extreme value statistics and, finally, the Letter is closed with a discussion.

\textbf{SDRG method.} - Our calculations are based on the SDRG method\cite{im}, which turned out to be very powerful for
the SR version of the model in Eq.(\ref{eq:H}). 
The critical behavior of that system in any finite dimension is controlled by a so called infinite-disorder fixed point (IDFP)\cite{danielreview}, where 
the dynamical scaling is extremely anisotropic, the time scale $\tau$,
and the length scale $\xi$ being related as $\ln\tau\sim \xi^{\psi}$ with an exponent $\psi$ depending weakly on $d$. 
The set of critical exponents has been calculated analytically in $d=1$\cite{fisher} and estimated numerically in $d=2,3,4$\cite{2d,2dki,ddRG}.
In the SDRG procedure, the terms
in the Hamiltonian with the largest parameter (either a coupling or a transverse field) are successively eliminated and effective parameters of the remaining system are calculated perturbatively\cite{mdh}. 
Decimating a large coupling $J_{ij}$, the spins $i$ and $j$ form a cluster, which has a moment $\tilde{\mu}=\mu_i+\mu_j$ and experiences an
effective transverse field $\tilde{h}={h_i h_j}/{J_{ij}}$. 
The coupling between this cluster and spin $k$ is
given by $\tilde{J}_{ij,k}={\rm max}(J_{ik},J_{jk})$.
Decimating a large transverse field $h_i$,
the actual spin is eliminated and new, effective
couplings $\tilde{J}_{jk}={\rm max}(J_{ji}J_{ik}/h_i,J_{jk})$
are generated between all pairs $(j,k)$ of its neighbors.
We used the efficient algorithm developed in Ref\cite{ddRG}
of the above SDRG scheme based on the 'maximum rule', which is asymptotically exact at an IDFP.

\begin{figure}[tb]
\begin{center}
\includegraphics[width=3.2in,angle=0]{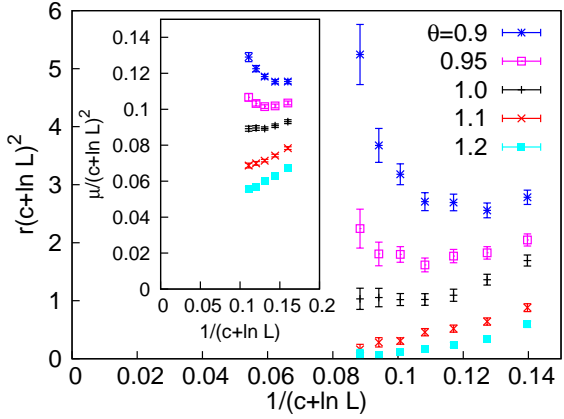}
\end{center}
\vskip -.5cm
\caption{
\label{fig_2} (Color online) Scaled decimation ratio $r_{\theta}(L)$ as a function of $L$ for different values of $\theta$.
The inset shows the scaled average magnetic moment.
}
\end{figure}

\begin{figure}[tb]
\begin{center}
\includegraphics[width=3.2in,angle=0]{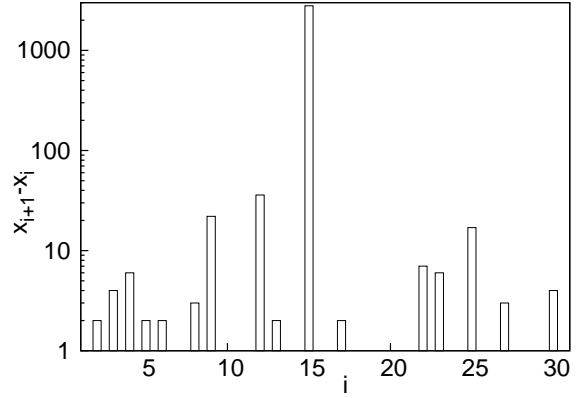}
\end{center}
\vskip -.5cm
\caption{
\label{fig_2a} Illustration of the structure of the largest non-decimated cluster consisting of $\mu=32$ spins in a sample of size $L=8192$.
The logarithm of the length of spacings between neighboring spins are given by the heights of the columns.
}
\end{figure}

\textbf{Numerical SDRG analysis.} - Focusing on the model in Eq.(\ref{eq:H}) in 1d, we started our investigations with a numerical
SDRG analysis. 
The parameters $b_{ij}$ and $h_i$ were uniformly distributed in the interval $(0,1]$  and $(0,h]$, respectively, and $\theta=\ln(h)$ was used as a control parameter of the transition. Other distributions have also been tried but they led to the same conclusions. 
The calculations were carried out for finite systems of sizes up to $L=4096$, in typically $40000$ realizations ($5000$ for the largest size), and with various decay exponents ($\alpha=2,3,4$) but, the conclusions being similar, we restrict ourselves to the presentation of numerical results for $\alpha=2$. 
\begin{figure}[tb]
\begin{center}
\includegraphics[width=3.2in,angle=0]{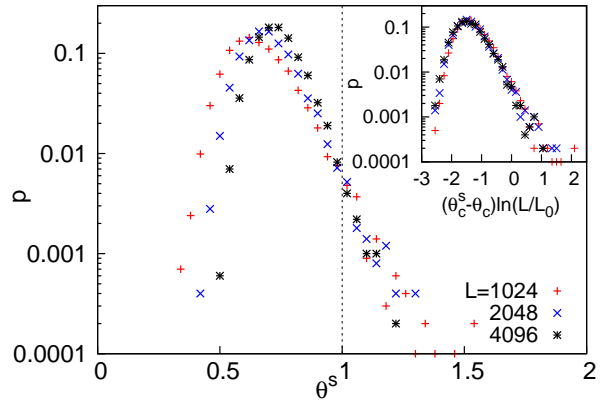}
\end{center}
\vskip -.5cm
\caption{
\label{fig_1} (Color online) Distributions of the pseudo-critical points, which cross each other for different $L$ at $\theta_c \approx 1$, indicated by a dotted line. The ratio of the accumulated distributions on two sides of $\theta_c$ is given by $r_{\theta_c}(L)$, see the text. The inset shows
the rescaled distributions.
}
\end{figure}
As a first step, for each random sample $s$, a pseudo-critical point $\theta_c^s$ has been determined\cite{2dki,ddRG},
with the definition, that for $\theta^s<\theta_c^s$ ($\theta^s>\theta_c^s$) the last decimated parameter is a coupling (transverse field).
Here, $\theta^s$ is the control parameter of the sample parameterizing the fields as $\ln h_i=\ln h_i^0+\theta^s$, where $h_i^0$ is uniformly distributed in $(0,1]$.
The distributions of $\theta_c^s$ for different sizes $L$ 
are shown in Fig.\ref{fig_1}. Using the scaling variable $(\theta_c^s-\theta_c)\ln(L/L_0)$ with $\theta_c \approx 1$, a
scaling collapse can be observed and, accordingly, the position of the maximum and the width of
the distributions scales as $ \sim 1/\ln L$. 
Note that for a conventional random fixed point the scaling combination is $(\theta_c^s-\theta_c)L^{1/\nu}$ \cite{domany,aharony,ddRG}.
The logarithmic scaling found here indicates an anomalous divergence of the correlation length of the form
\be
\xi \sim \exp\left(\rm const/|\theta-\theta_c|\right)\;.
\label{xi}
\ee
In addition to this, the ratio $r_{\theta}(L)$ of the frequency of the bond and field decimations, see Fig.\ref{fig_1}, is found to decrease to zero with increasing $L$ as 
\be
r_{\theta_c}(L)\sim 1/\ln^2(L),
\label{r_L}
\ee
as illustrated in Fig.\ref{fig_2}. 
The average magnetic moment $\mu(L)$ of the last remaining cluster is found to scale as
\be
\mu(L) \sim \ln^2 L\;
\label{mu}
\ee
in the critical point, as can be seen in the inset of Fig.\ref{fig_2}. 
This behavior contrasts again with the algebraic dependence in the SR model\cite{fisher,2d,2dki,ddRG} and means that the fractal dimension of
ferromagnetic clusters is formally zero.
The structure of the spin clusters is illustrated in Fig.\ref{fig_2a}.

\begin{figure}[tb]
\begin{center}
\includegraphics[width=3.2in,angle=0]{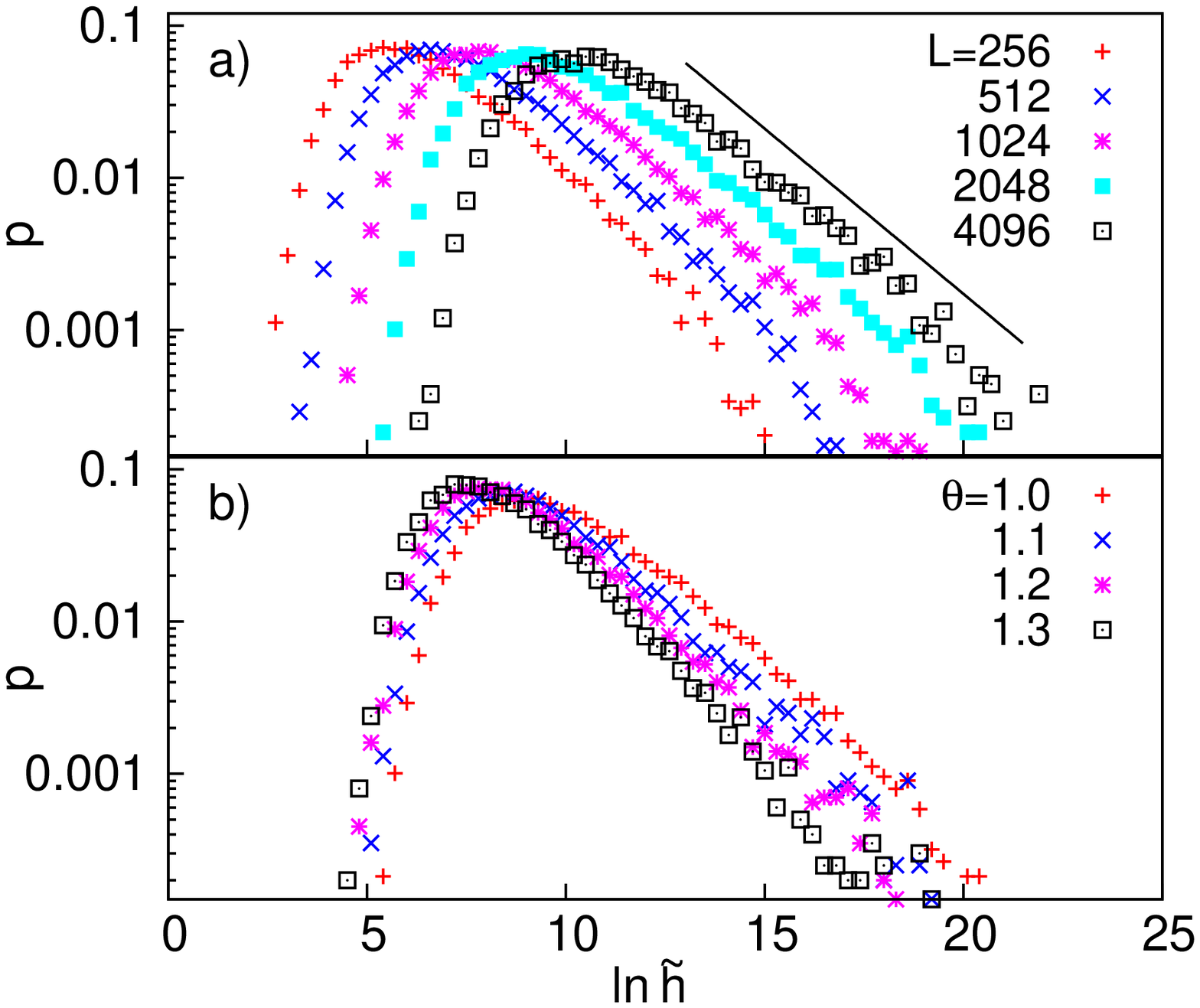}
\end{center}
\vskip -.5cm
\caption{
\label{fig_3} (Color online) a) Distributions of the logarithm of the last decimated transverse fields in the critical
point for different sizes. 
b) The same for $L=2048$ and for different $\theta$.
}
\end{figure}

The twice of the last decimated transverse field $\tilde{h}$ in a finite sample gives the lowest energy gap. In the critical point, the distribution of $\ln(\tilde{h})$ is found to shift with increasing $L$ by $z_c \ln L$ as can be seen in Fig.\ref{fig_3}a. Consequently, the correct finite-size scaling combination is $\tilde{h}L^{z_c}$, where $z_c$ is the critical dynamical exponent, which is extrapolated to be $z_c \simeq \alpha$.
For small values of $\tilde{h}$, the distributions have a power-law tail 
$g_L(\tilde{h}) \sim \tilde{h}^{1/z_c-1}$. 
This is consistent with the observation that, as bond decimations are rare, the last field is practically the smallest one out of $O(L)$ roughly independent fields and then extreme-value
statistics (EVS) \cite{galambos} explains the observed properties.
In the paramagnetic Griffiths phase, $\theta>\theta_c$, the distributions have the same form as for $\theta=\theta_c$, see Fig.\ref{fig_3}b, however, with a different, $\theta$-dependent dynamical exponent $z(\theta)<z_c$. We conclude that the critical dynamical exponent is finite, so the critical behavior is controlled by a \textit{strong-disorder fixed point} rather than an IDFP \footnote{Note that, in a strong-disorder fixed point, the asymptotic exactness of the SDRG results is not guaranteed \cite{im}.}. 

\textbf{Primary model.} - Analysing the SDRG procedure close to the fixed point in the paramagnetic phase and in the critical point,
we have a few observations, which can be used to simplify the SDRG scheme. First, almost always transverse fields are decimated; 
second, after a field decimation, the maximum rule leads almost always
to $\tilde{J}_{jk}=J_{jk}$; 
third, the extension $w_i$ of (non-decimated) clusters are typically much smaller than the distances between them. Let us now assume that the transverse fields are random, as before, but the couplings are non-random, i.e. $b_{ij}=b=1$, which, according to our numerical results, does not alter the universal properties.
Then, after decimating $h_i$, the effective coupling between nearest clusters $i-1$ and $i+1$ will always be smaller than the deleted ones, $J_{i-1,i}$ and 
 $J_{i,i+1}$. 
For the numerically found fixed-point distributions, we have almost always $\tilde{J}_{i-1,i+1}={J}_{i-1,i+1}$ and the renormalization rule of couplings between nearest clusters can be expressed in terms of the length variables as 
 $\tilde{J}_{i-1,i+1}^{-1/\alpha}=J_{i-1,i}^{-1/\alpha}+J_{i,i+1}^{-1/\alpha}+w_i$, where $w_i$ is neglected according to the third condition.
Using reduced variables 
$\zeta=\left(\frac{\Omega}{J}\right)^{1/\alpha}-1$ and $\beta=\frac{1}{\alpha}\ln\frac{\Omega}{h}$, the approximate renormalization rules are 
\be
\tilde\zeta=\zeta_{i-1,i}+\zeta_{i,i+1}+1\;
\label{zeta}
\ee
and
\be
\tilde\beta=\beta_i+\beta_{i+1}\;
\label{beta}
\ee
for field and bond decimation, respectively,
which define our \textit{primary model}.
Since, in the ferromagnetic phase, the effective couplings between remote clusters may be stronger than those between adjacent ones due to the large mass of clusters, this approach is justified in the paramagnetic phase and in the critical point only.
The evolution equations of the distributions $g_{\Gamma}(\beta)$ and $f_{\Gamma}(\zeta)$  under the increase of the logarithmic energy scale 
$\Gamma\equiv\frac{1}{\alpha}\ln\frac{\Omega_0}{\Omega}$, where $\Omega_0$ is the initial value of $\Omega$,  
are identical to those of the $1d$ disordered $O(2)$ quantum rotor model of granular superconductors \cite{akpr} with the grain charging energy 
$U_i$ and Josephson coupling $\mathcal{J}_{i,i+1}$  corresponding to 
$U_i\leftrightarrow J_{i,i+1}^{1/\alpha}$ and $\mathcal{J}_{i,i+1}\leftrightarrow h_{i}^{1/\alpha}$. Notice that this mapping
interchanges the on-site and interaction parameters of the models.

\textbf{Fixed-point solution of the primary model.} - The fixed-point solutions of the distribution functions
are exponentials\cite{akpr} $g_{\Gamma}(\beta)=g_0(\Gamma)e^{-g_0(\Gamma)\beta}$, $f_{\Gamma}(\zeta)=f_0(\Gamma)e^{-f_0(\Gamma)\zeta}$, where the scale factors $g_0(\Gamma)$ and  $f_0(\Gamma)$ obey 
the differential equations 
\be
\frac{dg_0(\Gamma)}{d\Gamma}=-f_0(\Gamma)g_0(\Gamma), \quad 
\frac{df_0(\Gamma)}{d\Gamma}=f_0(\Gamma)(1-g_0(\Gamma)),
\ee
and have the $\Gamma \to \infty$ limits 
$f_0(\Gamma)\to 0$ and $g_0(\Gamma) \to 1+a$ in the paramagnetic phase $a>0$ and in the critical point $a=0$. 
Close to the critical point ($0\le a\ll 1$), the solutions can be written in the compact form valid in leading order in $\Gamma$:
\beqn
g_0(\Gamma)&\simeq&1+a{\rm coth}[(\Gamma+C)a/2], \nonumber \\
f_0(\Gamma)&\simeq&\frac{a^2}{2{\rm sinh}^2[(\Gamma+C)a/2]}\;,
\label{gf}
\eeqn
with a constant of integration $C$.
The fraction of non-decimated sites $n$ satisfies the differential equation $\frac{dn}{d\Gamma}=-n(g_0+f_0)$, from the solution
of which we obtain a relationship between the energy cut-off, $\Omega$ and the length scale, $l=1/n$ in the form 
\be 
l \simeq e^{\Gamma}a^{-2}{\rm sinh}^2[(\Gamma+C)a/2] \sim \left(\frac{\Omega_0}{\Omega}\right)^{\frac{1+a}{\alpha}}\;,
\label{dyn}
\ee
with an additional factor $\ln^2(\Omega/\Omega_0)$ in the last expression for $a=0$.
Thus, the dynamical exponent, $z=\alpha/(1+a)$, is a continuous function of $a$ and it is maximal, but finite at the critical point:
$z_c=\alpha$. 
The limit distribution of the transverse fields for $\Gamma \to \infty$ follows a power law $g(h) \sim h^{1/z-1}$ in agreement with the numerical SDRG results
shown in Fig. \ref{fig_3}.

Using Eq.(\ref{dyn}) and that, according to Eq.(\ref{gf}), 
the appropriate scaling combination is $\Gamma a=C'$ in the vicinity of the critical point,
we have for the characteristic length scale
$\xi\sim\exp(\Gamma)\sim\exp(C'/a)$. This is in agreement with the 
numerical finding in Eq.(\ref{xi}) with $\theta-\theta_c \sim a$. 
The decimation ratio in the primary model 
$r(\Gamma)={f_0}/{g_0}$ scales in the critical point with the system size as $r(L)\simeq 2 \ln^{-2} (L/L_0)$. This agrees again with the behavior found numerically (see Fig. \ref{fig_2}), although the prefactor appears to be different there owing to strong corrections to the leading behavior for moderate $L$. 
We have also calculated the average mass $\mu(L)$ of the last remaining 
clusters, which is related to the average spontaneous magnetization as $m(L)=\mu(L)/L$. 
The analogous quantity in the quantum rotor model is out of interest and has not been calculated.
Adapting the way of calculation of $\mu(L)$ in the SR model \cite{igloi02} to the present case, we obtain $\mu(L) \sim \ln^2L$ in the critical point, in agreement with the numerical SDRG results shown in the inset of Fig. \ref{fig_2}. 

Finally, we consider the entanglement entropy $S_L$ of a finite block of size $L$ in an infinite system. 
In the SDRG approach, $S_L$ is given by the number of decimated bonds that connect spins inside the block with those outside \cite{refael_moore}.
In the critical primary model, $S_L$ can be calculated 
by making use that the length of a bond decimated at $\Omega$ is $\ell=\Omega^{-1/\alpha}$, which yields for the mean number of decimated bonds with length $\ell$ per unit length of the system 
$B(\ell)d\ell\sim\ell^{-2}\ln^{-4}\ell \mathrm{d}\ell$. The asymptotic size-dependence of the entropy is given by $S_L\sim \int^L\ell B(\ell)\mathrm{d}\ell=S_{\infty}+O(\ln^{-3}L)$.
Thus, $S_L$ saturates in the limit $L\to\infty$
as opposed to the logarithmic divergence $S_L\sim \ln L$ characteristic of  both disordered \cite{refael_moore} and pure\cite{vidal}
$1d$ critical SR systems.
The boundedness of the entanglement entropy is related to the extremely dilute structure of critical clusters, which is illustrated in Fig. \ref{fig_2a}. 
The numerical results obtained by the SDRG method are compatible with the theoretical form, including the logarithmic correction term.

\textbf{Interpretation through EVS.} - As can be seen, the primary model has proved to be very useful to obtain the
asymptotic scaling behavior of the $1d$ LR model.
Some results can also be inferred by EVS in a heuristic way, as follows.
Let us have a finite chain of
length $L$, which is renormalized to a cluster of $\mu$ spins. According to the decimation rules, its
effective field is expressed in terms of the original parameters as
 $\tilde{h} \sim \prod_{i=1}^{\mu} h_i/ \prod_{i=1}^{\mu-1} J_i$,
where $J_i=b_i r_i^{-\alpha}$ and, now, $b_i$ is not necessarily homogeneous.
Using that the limit distribution of the fields is $g(h)\sim h^{-1+(1+a)/\alpha}$, $h_i$, being the smallest out of $r_i$ variables,
is given according to EVS as $h_i \simeq \kappa_i r_i^{-\alpha/(1+a)}$, where the random numbers $\kappa_i$ follow Fr\'echet
statistics\cite{galambos}: $P(\kappa)=\alpha^{-1}\kappa^{1/\alpha-1} \exp(-\kappa^{1/\alpha})$. 
The asymptotic behavior of the above expression of $\tilde h$ is different  
for $\overline{\ln h}>\overline{\ln J}$ and $\overline{\ln h}<\overline{\ln J}$, where the overbar denotes an average over disorder,
yielding the criticality condition $a=0$ and $\overline{\ln b}=\overline{\ln \kappa}$. 
In the critical point, we have thus $ \prod_{i=1}^{\mu} \kappa_i/ \prod_{i=1}^{\mu-1} b_i \sim \exp(-c \mu^{1/2})$ from the central
limit theorem and, on the other hand, $\tilde{h} \sim L^{-\alpha}$, which implies $\mu \sim \ln^2 L$.

In the paramagnetic phase with $0< a \ll 1$ the correlation length, $\xi(a)$, is defined by the length of the longest decimated bond, $r_l$,
the strength of which satisfies the relation:
$J_l/h_l \sim (b_l r_l^{-\alpha})/(r_l^{-\alpha/(1+a)}\kappa_l) \sim r_l^{-\alpha a}/\kappa_l > 1$.
Thus, the smallest value of
the decimated transverse fields has a parameter: $\kappa_l < \xi^{-\alpha a}$.
We have then ${\rm Prob}(\kappa_l < \xi^{-\alpha a})={\cal O}(1)$, which, using that
the variable $\kappa_l$ follows Fr\'echet statistics, can be written as
\be
\int_0^{\xi^{-\alpha a}} P(\kappa) {\rm d} \kappa=1-e^{-\xi^{-a}}
\sim \xi^{-a} = e^{-C'}={\cal O}(1).
\ee
We obtain thus for the correlation length $\xi \sim \exp(C'/a)$, in agreement with Eq.(\ref{xi}).

\textbf{Beyond maximum rule.} - Both the numerical and analytical approaches applied so far were based on the maximum rule. 
In the remainig part of this work, we go beyond this limitation by taking into account interactions between all spin pairs of adjacent clusters. 
Then, the total coupling $J_{i,i+1}$ will be roughly $\mu_i\mu_{i+1}$ times the bare coupling of a pair of spins.
Defining now the variable $\beta_i$ as $\beta_i=\frac{1}{\alpha}\ln\frac{\Omega \mu_i^2}{4h_i}$, it will be additive under the decimation of a bond $J_{i,i+1}$, as before,
provided that $\Omega$ is defined as $\Omega=J_{i,i+1}/\tilde \mu_{i,i+1}^2$ with $\tilde \mu_{i,i+1}=\frac{\mu_i\mu_{i+1}}{\mu_i+\mu_{i+1}}$.
Similarly, the variable 
$\zeta_{i,i+1}=\left(\frac{\Omega\tilde \mu_{i,i+1}^2}{J_{i,i+1}}\right)^{1/\alpha}-1$ with $\Omega=\frac{4h_i}{\mu_i^2}$
would transform as $\zeta_{i,i+2}=\zeta_{i,i+1}+\zeta_{i+1,i+2}+1$ under the decimation of a field $h_i$ 
as in the primary model, if all the clusters before the decimation had equal masses. Although this is not strictly the case,
the narrow distribution of cluster masses $\rho(\mu)\sim\exp(-c\sqrt{\mu})$ leads to the appearance of $O(1)$ random multiplicative
factors in the decimation rules, which do not modify the asymptotical properties. Therefore all the results obtained for the
primary model are expected to be valid here, but for the reduced couplings $\frac{J_{i,i+1}}{\tilde \mu_{i,i+1}^2}$
and $\frac{4h_i}{\mu_i^2}$. 
This conclusion has been confirmed numerically by an improved version of the primary model, 
in which couplings between all pairs of spins in adjacent clusters were taken into account.

\textbf{Discussion.}
- In summary, we have studied the low-energy properties of the random transverse-field Ising model with long-range interactions in $1d$ 
by a numerical SDRG method and by the analysis of a simplified scheme, which is
expected to contain the relevant ingredients of the SDRG procedure. 
These revealed an unusual critical behavior controlled by a strong disorder fixed point with a finite dynamical exponent $z_c=\alpha$. 
The correlation length was found to diverge exponentially,
the magnetization obeys a logarithmic finite-size scaling law and 
the entanglement entropy satisfies the area law even in the critical point.
Except of the dynamical exponent, other properties of the phase transition are found to be independent of the value of $\alpha>1$.
Thus the scenario with $\alpha$-dependent critical regimes, as generally observed in other problems with LR interactions, does
not hold for the LR random transverse-field Ising chain. 
Knowing that, in the SR model, the strength of the effective couplings decreases with the renormalized length scale typically as $J\sim\exp(-Cl^{1/2})$ \cite{fisher}, the universality class of the SR model is expected to be recovered 
by a faster decay of the interactions of the form $J(r)\sim\exp(-Cr^{\omega})$
with $\omega>1/2$. 
Here we note, that although the methods applied in this work provide a coherent picture of the low-energy behavior of the model, it is not guaranteed that the fixed point obtained here describes the critical behavior of the model for \emph{any weak} disorder. The question of possible existence of a different fixed point in this regime is out of the scope of the present approach and needs an alternative investigation. 

The results obtained in this work may be relevant in general for systems where the critical behavior is determined by the interplay of (quantum) fluctuations, disorder and long-range interactions. 
We mention models with a discrete order parameter such 
as the $q$-state quantum Potts model\cite{senthil} and the quantum Ashkin-Teller model\cite{carlon-clock-at};
these are expected to have the same fixed point as obtained here. 
In the field of non-equilibrium processes, we can mention the contact process in a random environment\cite{cp}.
The SDRG decimation rules for this model are very much similar to those of the random transverse-field Ising model and for
SR interactions both models have the same IDFP, at least for strong enough disorder\cite{hiv}. This correspondence is found to be valid for generalized small-world networks as well\cite{jk}, where, instead of the strength $J_{ij}$, the probability of the long-range interactions decays with a power, $\alpha=2$.
The two models are expected to be in the same universality class also for
LR interactions, which has been checked by Monte Carlo simulations\cite{inprep}, providing thereby an independent confirmation of the validity of our SDRG approach.
Considering models with a continuous symmetry, such as the LR random antiferromagnetic Heisenberg chain, the dynamical exponent is still expected
to be $z_c=\alpha$, as for the LR random transverse-field Ising chain.

An interesting question is the behavior of the model in higher dimensions.
Preliminary numerical SDRG analyses in $2d$ indicate that the critical behavior is similar to that found in $1d$.
The critical properties are controlled by a strong-disorder fixed point with a dynamical exponent $z_c=\alpha$,
the length scale diverges exponentially and the ferromagnetic clusters have a zero fractal dimension. 
This may be due to the fact that dimensionality plays a less important role in the presence of long-range interactions.
The detailed study of this question is deferred for future work.

\begin{acknowledgments}
This work was supported by the National Research Fund 
under grant no. K75324, K77629, and K109577; by the J\'anos Bolyai Research Scholarship of the
Hungarian Academy of Sciences (RJ), and partially supported by the European Union and the
European Social Fund through project FuturICT.hu (grant no.:TAMOP-4.2.2.C-11/1/KONV-2012-0013). 
The research of IAK was supported by the European Union and the State of Hungary, co-financed by the European Social Fund in the framework
of T\'AMOP 4.2.4. A/2-11-1-2012-0001 'National Excellence Program'.
\end{acknowledgments}

\end{document}